# Anomalous decoupling dynamics in glycerol and its nanocolloid with silver nanoparticles

[1]Szymon Starzonek, [1,2]Sylwester J. Rzoska, [2]A. Drozd-Rzoska, [1]Sebastian Pawlus, [3]Julio Cesar Martinez-Garcia and [4]Ludmila Kistersky.

[1]ŚMCEBI & Institute of Physics, University of Silesia, ul. 75 Pułku Piechoty 1, 41-500 Chorzów, Poland

[2]Institute of High Pressure Physics, Polish Academy of Sciences, ul. Sokołowska 27/39, Warsaw 01-142, Poland.

[3]University of Berne, Freiestrasse 3, Berne CH-3012, Switzerland.

[4]V. Bakul Institute for Superhard Materials of the National Academy of Superhard materials NASU, Avtozavodskaya Str.2, 04074 Kiev, Ukraine

# Abstract

Orientational – translational (*T&O*) decouplings in ultraviscous glass forming glycerol and its nanocolloid based on Ag nanoparticles, as the function of temperature and pressure up to challenging $P > 1.5 GPa$, are discussed. The analysis is focused on the fractional Debye-Stokes-Einstein relation (*FDSE*) $\sigma(T,P)[\tau(T,P)]^S = const$, where $\tau$ and $\sigma$ are for the structural relaxation time and electric conductivity, respectively. For temperature tests in glycerol the clear evidence for the "almost coupled" exponent $S \approx 1$ was obtained. For the supercooled nanocolloid the crossover to the decoupled domain associated with $S \approx 0.91$ was noted. In superpressed glycerol and the nanocolloid the clear cross over from the domain described by the exponent $S \approx 1$ to $S < 1$ appeared on the GPa domain. For the nanocolloid it is associated with particularly strong decoupling, related to $S \approx 0.5$, and occurs at well-defined pressure. This can suggest a possible fluid-fluid transition. Questions related to the validity of the Debye-Stokes $\tau \propto \eta/T$ and Maxwell $\tau \propto \eta$ relations ($\eta$ is for viscosity) are also addressed. Results obtained do not support the recently suggested universality of the fractional exponent describing *T&O* decoupling.

**Introduction**

"Universal" features of the previtreous dynamics on approaching the glass temperature ($T_g$) constitute on the most attractive and challenging problems of the modern condensed and soft matter physics.[1-3] This indicates a possible "unified" description of notably different systems, what is one of basic aims of physics.[4] They are for instance:[3] (*i*) super-Arrhenius (SA) evolution of relaxation time ($\tau$), viscosity ($\eta$) or related dynamic properties, (*ii*) non-Debye broadening of the distribution of relaxation times, (*iii*) decoupling of the translational and orientational degrees of freedom, (*iv*) the emergence of the secondary (*beta*) relaxation,...

Recently, these "functional" similarities have been supplemented by a set of well-defined values of empirical parameters associated with them. For the SA dynamics apart from the *fragility* as the metric of the distortion from the SA dynamics, the new and local symmetry related parameter *n* has been introduced. It is linked to three characteristic values $n \approx 3/2$ (orientational, uniaxial symmetry), $n \approx 0.2$ (positional symmetry) and $n \approx 1$ (negligible local symmetry). The latter is associated, for instance, with glycerol which is considered as one of the most "classical" glass forming liquids. The Vogel–Fulcher–Tammann (VFT), which is consider as the basic tool for portraying $\tau(T)$ or $\eta(T)$ SA behavior, is optimal only for glass formers where $n \approx 1$.[5,6] It was also discovered that the coefficient describing the shape of the high frequency part of the relaxation time distribution loss curve $\varepsilon''(f > f_{peak}) \propto f^{-ab}$: $ab \to 1/2$ for $T \to T_g$, for low molecular weight liquids,[7,8] liquid crystals and plastic crystals[9] and also for isothermal, pressure approaching the glass transition[9,10]. For the translational-orientational decoupling Mallamace et al.[11] discovered that the "parameter universality" can be also the case of the translational-orientational in relations linking diffusivity (*D*), structural relaxation time ($\tau$) and viscosity ($\eta$), namely:

$$D/T = A\eta^{-\zeta} \qquad \text{and} \qquad D/T = A\tau^{-\zeta} \qquad (1)$$

with the same universal value of the fractional exponent $\zeta \approx 0.85$, empirically evidenced for few tens of liquid glass formers, including glycerol.[11]

The appearance of mentioned "universalities" is explained by invoking dynamic heterogeneities, related to the emergence of transient spatially and temporally separated domains on approaching the glass transition. This issue is particularly stressed for the translational – orientational ($T\&O$) decoupling.[3,11 and refs. therein.]
All these may resemble the situation in the physics of critical phenomena before the ultimate solving of this great conceptual puzzle,[12] and suggest that the long awaiting fundamental breakthrough in the glass transition physics is approaching. Consequently, studies focusing on properties recalled above can be particularly important.

This report focuses on the translational – orientational ($T\&O$) decoupling in ultraviscous glycerol, probably the most "classical" glass forming liquid.[3] Studies were carried out via the comparison of electric conductivity and relaxation time, still poorly evidenced for glycerol. They also recall basic questions of the validity of $\tau \propto \eta$ vs. $\tau \propto \eta/T$ relations, the universality indicated by Mallamace et al.[11] (eq. (1)) and show the notable impact of two exogenic factors: (*i*) silver (Ag) nanoparticles, yielding a stable nanocolloid without any additional component, and (*ii*) the hydrostatic pressures up to extreme $P \approx 1 GPa$

**Translational-orientational decoupling**

As the onset of the ultraviscous domain the dynamic crossover temperature $T_B(\tau \sim 10^{-7\pm 1}) \approx 1.2 T_g$ is considered.[3, 13, 14] The temperature $T_B$ is linked the emergence of the secondary relaxation, mode-coupling "critical" temperature or the appearance of the decoupling between orientational and translational degrees of freedom.[3, 11, 13-16] Generally, the dynamic crossover is associated with the crossover from the ergodic to non-ergodic domain on approaching the glass transition and the appearance and growth of cooperatively rearranged regions / dynamic heterogeneities.[3] The translationa – orientational (*T&O*) decoupling is manifested by the emergence of the behavior described by fractional Debye-Stokes-Einstein (*FDSE*) relation, instead of basic *DSE* dependences obeying in "normal" liquids:[3, 15-19]

$$\frac{D}{T} = \frac{Ck_B}{a}\eta^{-1} \quad (2), \qquad \tau = \frac{v\eta}{k_B T} \quad (3), \qquad \frac{D}{T} = \frac{Cv}{aT}\tau^{-1} \quad (4)$$

where $v = A'V$ and $a$ denote the constant depending on the form of the diffusing molecule, $V$ is for the molecular volume, $D$ is the diffusion coefficient, $C$ is the constant distinguishing rotational and translational diffusivity, $\eta$ is for shear viscosity and $\tau$ is for the primary (*alpha*) relaxation time determined within BDS tests from the peak frequency of the loss curve via $\tau = 1/(2\pi f_{peak})$; $k$ is the Boltzmann constant and $T$ is for temperature.

The DS eq. (3) indicates that $\tau \propto \eta/T$, but there is also the alternative approach resulting from the Maxwell relation[18] $\tau = G_\infty \eta$. Assuming that in the ultraviscous domain the instantaneous shear modulus $G_\infty = const$ it gives $\tau \propto \eta$. It is worth recalling that in the Maxwell relation $\tau$ denotes the stress relaxation time and there are no clear experimental evidence that the structural and stress relaxation times are interchangeable.[3, 18]

Linking above dependences with the Nernst-Einstein relation $D = k_B T\sigma/ne^2$,[17] where $n$ is the number of electric charges/carriers and $\sigma$ denotes the electric conductivity, one obtains:

$$\sigma\tau = \frac{ne^2 Cv}{k_B T} \quad \text{, i.e.} \quad \frac{\sigma\tau}{T} = const \quad (5) \quad \text{or} \quad \sigma\tau = \frac{CG_\infty ne^2}{a}, \text{ i.e. } \sigma\tau \approx const \quad (6)$$

Then, tests focused on the behavior of $\sigma$ vs. $\tau$ dependence can offer a tool for distinguishing between tests offered by DS eq (3) (via eq. (5)) and the Maxwell relation via eq.(6) scenarios. It is notable that experimental $\sigma(T)$ and $\tau(T)$ dependences can be obtained with high resolution, from the same broad band dielectric spectroscopy (BDS) scan covering even more than 12 decades in time and frequency. This reduces biasing experimental artifacts which can appear when using experimental data from different experiments. In ultraviscous liquids, for $T_B < T < T_g$, the "fractional" form of DSE (FDSE) relations emerges. This is the case of eq. (1) as well as "fractional" forms of eqs. (5) or (6):[14-16, 19-39]

$$\sigma \tau^S = const \qquad (7)$$

where the FDSE exponent $S < 1$.

Recalling the "universal" value of the exponent $\zeta \approx 0.85$ in eq. (1), suggested empirically by Mallamace et al.[11], and subsequently the SE eq. (2), the Maxwell equation and the Nernst-Einstein relation, one obtains:

$$\sigma \tau^\zeta = \frac{CG_\infty^{-\zeta} n e^2}{a}, \quad \text{i.e. the fractional exponent} \quad S = \zeta \qquad (8)$$

The empirical universality of the fractional exponent $\zeta$ linking $D \& \eta$ and $D \& \tau$, proposed in ref.[11], may extend also for the FDSE dependence linking $\sigma \& \tau$: $S = \zeta \approx 0.85$.

Regarding the experimental evidence focusing on mutual changes of the structural relaxation time and electric conductivity in the ultraviscous domain, only in ref.[25] for polyvinylmethylether the validity of relation $\sigma \tau^S / T = const$ was shown. The lack of the temperature dependence, i.e. the experimental validity of eq. (7) for the vast majority of experiments carried out so far, is explained via the statement that in the tested range of temperatures in ultraviscous liquids the change of temperature is small and negligible.[19-21] In the opinion of the authors this statement poorly coincides with the fact that the ultraviscous domain can reach even $\Delta T \approx 100 K$. As shown above (eqs. (5) and (6)) the response to the question regarding the validity of $\sigma \tau / T = const$ or $\sigma \tau = const$ can be associated with impacts of the Maxwell relation or the DS equation in ultraviscous liquids.

It notable that in the high temperature dynamic domain for $T > T_B$, the T&O coupling is restored and exponents $S = 1$ and $\zeta = 1$.[3, 11, 19-21] It is worth recalling that above $T_B$ the evolution of $\tau(T)$ or $\eta(T)$ remains non-Arrhenius type but it is portrayed via the mode-coupling theory (MCT) based "critical-like" relations $\tau(T), \eta(T) \propto (T - T_C)^{-\phi}$, where $T_C + 100K > T > T_C + 20K$ and $T_C = T_B$. Only for higher temperatures the basic Arrhenius behavior, $\tau(T), \eta(T) \propto \exp(\Delta E_a / RT)$ with $\Delta E_a = const$, takes place.[3]

Worth noting is the evidence questioning the ability of the FDSE eq. (7) for portraying experimental data in ultraviscous liquids. Notable distortions were observed for ultraviscous electrolyts.[29] For supercooled propanol the decrease from $S \approx 1$ to $S \approx 0.7$, starting ca. 30 K above $T_g$, was reported.[31] Recently, Shi et al.[39] indicated problems of DSE and FDSE behavior basing on the theoretical analysis of numerical model systems.

**The impact of high pressure**

The super-Arrhenius (SA) evolutions of viscosity $\eta$, primary relaxation time $\tau$, diffusion coefficient $D$ or DC electric conductivity $\sigma$ are considered as the key hallmark of the previtrous dynamics in glass forming, ultraviscous liquids. This SA behavior is mostly portrayed via the Vogel-Fulcher-Tammann (VFT) equation:[3, 15]

$$x(T) = x_0 \exp\left(\frac{\Delta E_a(T)}{RT}\right) = x_0 \exp\left(\frac{D_T T_0}{T - T_0}\right) \quad (9)$$

where the magnitude $x$ is related to $\tau$ or $\eta$, $D$ or $1/\sigma$. The first, left-hand, equation is for the general SA behavior with the temperature dependent apparent activation energy $\Delta E_a(T)$ The subsequent equation has the VFT form in which $D_T$ denotes the fragility strength coefficient and $T_0 < T_g$ is the VFT estimation of the ideal glass temperature.

In the case of glycerol it is located ca. 30 K below the glass temperature $T_g$ empirically related to $\tau(T_g) = 100s$. The basic Arrhenius equation is restored for $\Delta E_a(T) = \Delta E_a = const$.

In the last years the adequacy of the VFT equation for portraying dynamics in ultraviscous, liquids has been strongly criticized, leading to the focus on relations avoiding a finite temperature divergence (like $T_0$ in eq. (1)).[40-43] Consequently, the existence of the ideal glass temperature (Kauzmann temperature), as well as a hypothetical phase transition hidden below $T_g$, have been put in doubt.[40, 44] However, a very recent discussion based on the analysis of the activation energy temperature index evolution showed that the optimal parameterization of $\tau(T)$ or $\eta(T)$ is clearly associated with the finite temperature divergence.[5,6] It also appeared that the VFT equation constitute an optimal tool of portrayal only for a narrow group of glass formers, where the new local symmetry related parameter $n \approx 1$.[6] Otherwise, the VFT equation can be considered as a practical, but only effective, tool of parameterization. Fortunately, glycerol belongs to the limited group where the VFT equation remains valid.[6]

For the isothermal, pressure related approaching the glass transition point $(T_g, P_g)$ one can use the parallel of the VFT equation:[45]

$$\tau(T) = \tau_0^P \exp\left(\frac{\Delta P V(P)}{RT}\right) = \tau_0 \exp\left(\frac{\Delta P D_P}{P_0 - P}\right) \tag{10}$$

where $V(P)$ denotes the apparent free volume energy related to the mechanisms detected by the given monitoring method ($\tau$ or $\eta$, $D$ or $1/\sigma$). $\Delta P = P - P_{Sp}$, where the latter is for the absolute stability limit hidden in the negative pressure domain. $D_P$ denotes the pressure related fragility strength coefficient.

The introduction of the negative pressures domain[46] in eq. (10) removed the essential inconsistency associated the arbitrariness of the value of prefactor $\tau_0^P$ for previously used equations $\tau(T) = \tau_o^P \exp(A/(P_0 - P))$[47, 48] and $\tau(T) = \tau_o^P \exp(D_P P/(P_0 - P))$[49]. It also takes into account that the solid state does not terminates at $P = 0$, but extents down to the absolute stability limit spinodal $(T_{Sp}, P_{Sp})$ hidden in the negative pressures domain.[45,46]

The pressure counterpart of the simple Arrhenius equation was proposed by Barus, namely $\eta(P) = \eta_0^P \exp(CP)$.[50] It can be restored from eq. (10) assuming $V(P) = V = C = const$ and $P_{Sp} = 0$. In the opinion more optimal can be the "upgraded" Barus equation taking, which can be concluded from eq. (2) and takes into account the real extend of the solid phase, namely $\tau(P) = \tau_0^P \exp(C'\Delta P)$.

The comparison of eqs. (9) and (10) indicates that the temperature evolution of dynamic properties is related to apparent activation energy and the pressure behavior to the apparent activation volume. This difference causes that compression and heating can have different impact different impact on previtreous phenomena, thus leading to their enhancing, splitting or even emerging.

Regarding the translational – orientational decoupling the existing evidence shows that on cooling and pressuring towards the glass transition:[21, 23-27]

$$\sigma(T,P)\tau(T,P)^S = const \qquad (11)$$

with the same value of the FDSE exponent *S* for temperature and pressure paths. The theoretical support for. (11) was proposed by Johari & Andersson.[28] However, systematic deviations from the FDSE behavior under compression were noted by Johari and Andersson in pressurized acetomitophen – aspirin mixture for $10^{-6} < \tau < 10^{-3} s$.[29]

**Experimental**

Fluid nanocomposite/nacolloidal mixture with the concentration reaching 180 ppm of 30 nm Ag (silver) nanoparticles in glycerol was prepared in the Institute of Superhard Materials in Kiev, Ukraine. It is notable that no additional chemicals or surfactants were needed to stabilize the nanocolloid and avoiding the sedimentation. Ag nanoparticles were synthesized via the localized ion-plasma sputtering and immediate implantation of freshly created nanoparticles to the carrier liquid in vacuum what allows to produce highly concentrated stable dispersions of ultra clean .metals nanoparticles in various carrier liquids.[51] The size distribution of nanoparticle, averaged at ~ 30 *nm* was below 2 %. At least 1 year stability for the nanocolloid was reached. All these results is comparable with the ultraclean solutions produced by laser ablation in liquids (LAL) but the combined ion-plasma sputtering demonstrate much higher productivity and cost effectiveness.[52]

Dynamics of the pressurized glycerol and Ag-glycerol nanocolloid were tested via the piston-based pressure set-up, described in ref.[53]. The gap of the flat parallel measurement capacitor was equal to 0.2 mm. The BDS spectrum, up to ca. 1.6 GPa, was monitored via the BDS Alpha Novocontrol spectrometer, giving 6 numbers resolution. This report focuses on the pressure evolution of DC conductivity $\sigma$ and the primary relaxation time $\tau$. The first one was estimated directly from the low frequency part of BDS spectra and the relaxation time was determined from the peak the primary loss curve via $\tau = 1/2\pi f_{peak}$ condition.[15] Pressure studies were carried out for $T \approx 258K$ isotherms, the lowest possible for the given experimental set up. It was selected to cover the broadest time-scale possible.

**Results and Discussion**

There is an extensive evidence regarding broad band dielectric spectroscopy (BDS) studies in ultraviscous glycerol. It indicates the clearly SA behavior, well portrayed by the VFT eq. (9) with the VFT singular temperature $T_0 \approx 160K$, for $T_g \approx 191K$.[3, 6] As shown in ref.[6] for glycerol the temperature $T_0$ coincides with the Kauzmann temperature ($T_K$), i.e. for glycerol there is a fair agreement between dynamic and thermodynamic estimation of the ideal glass temperature.

Glycerol belongs to the group of relatively "strong" glass formers, with the SA dynamics characterized by the fragility $m = \left[d\log_{10}\tau(T)/d(T_g/T)\right]_{T \to T_g} \approx 78$.[6] Values of this coefficient ranges from $m > 170$ for strongly SA behavior ("very fragile" glass formers), to $m \approx 16$ for the non-SA, simply Arrhenius behavior.[3, 5, 6] Glass formers with near-Arrhenius dynamics are known as "strong" glass formers.[3]

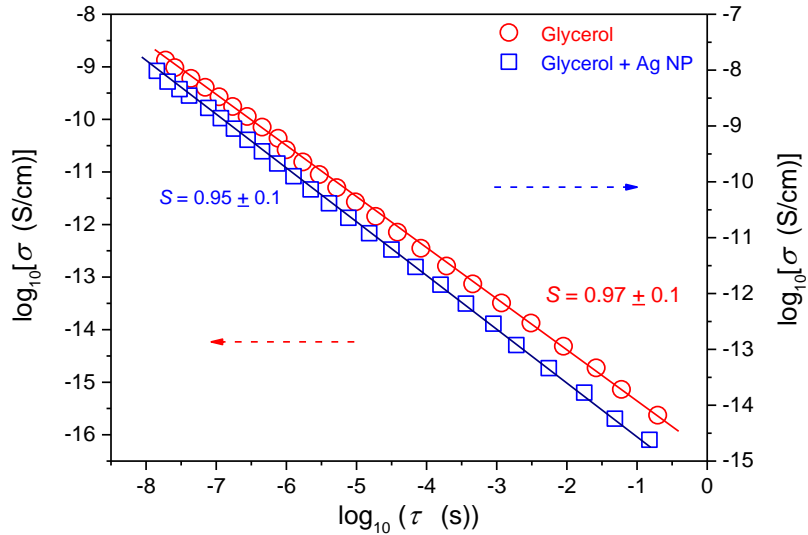

**FIG. 1**  *Temperature test of DSE law in pure glycerol and glycerol + AgNP composite at the pressure P = 0.1 MPa. The right scale is for the nanocolloid and the left one for glycerol. Slopes of lines, determining FDSE exponent, are also given*

Results presented in Fig. 1 clearly shows that for ultraviscous glycerol the non-Debye and super-Arrhenius dynamics is assisted by the FDSE eq. (7) with the "non-FDSE" exponent $S \approx 0.95$. Such behavior takes place without notable distortions, as shown using the distortions-sensitive analysis in Fig. 2.

For the nanocolloid, the basic plot used for FDSE analysis, $\log_{10}\sigma$ vs. $\log_{10}\tau$,[16, 19-31] indicate that $S \approx 1$. However, the distortions-sensitive analysis presented in Fig. 2 revealed a clear crossover/transition from the domain described by the exponent $S=1$ to $S=0.91$ ($\pm 0.01$).

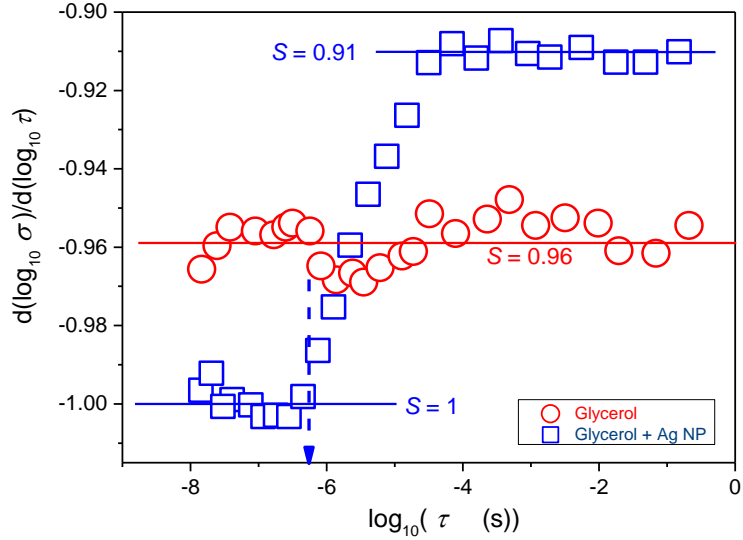

**Fig. 2** *The derivative of FDSE experimental data from Fig. 1, focused on the distotions-sensitive test of the validity of eqs. (7). The numerical filtering procedure as in refs.[5,6] was implemented. Solid horizontal lines show values of related FDSE exponents. The dashed arrow indicates the onset of the FDSE domain in the nanocolloid.*

The pressure evolution of the glass temperature in glycerol can be well portrayed by:[54, 55]

$$T_g(P) = T_{ref.}\left(1 + \frac{\Delta P}{\pi + P_{ref.}}\right)^{1/b} \times \exp\left(\frac{\Delta P}{c}\right), \text{ for } \Delta P = P - P_{ref.}, \qquad (12)$$

where for glycerol $-\pi \approx -0.25 GPa$ is the extrapolated, negative pressure for which $T_g(P \to -\pi) \to 0$ and the exponent $b \approx 2.4$. For the damping term $c \approx 2.0 GPa$. Regarding $(P_{ref.}, T_{ref.})$ one can assume arbitrary value along the melting curve.

Its notable feature is the hypothetical maximum at $T_g^{max} \approx 301K$ and $P_g^{max} \approx 6.1GPa$. For tests on the pressure of isothermal, pressure dynamics the isotherm $T = 258K$, related to $P_g \approx 1.95GPa$ was used.

It is notable that existing references for isothermally pressured glycerol are puzzling. Pronin et al. [56] carried isothermal BDS tests at ca. 290 K from ca. 2 up to 6 GPa and obtained a fair agreement with the VFT-type equation $\tau(P) = \tau_0 \exp(D_P P/(P_0 - P))$,[47] for lower pressures $P < 1GPa$). Earlier, Forsmann et al.[57] reported the simple Barus-type[50] behavior $\tau(P) = \tau_0 \exp(AP)$ up to ca. 3 GPa at room temperature. On the other hand, BDS test up to ca. 0.4 GPa by Paluch et al.[49] indicated weakly SA/SB behavior. Reiser and Kasper[58] tested the temperature evolution of $\tau(T)$ in glycerol for 6 isobars, up to $P$ = 600 MPa, and obtained a clearly overlapped SA behavior. Worth recalling are also studies by Herbst et al.[59], where using the centrifugal method the negative pressures domain $P < 0$ were entered. This result experimentally showed that pressure $P = 0$ or $P = 0.1MPa$ do not constitute a specific reference or terminal for ultraviscous liquids.

Results presented in Fig. 3 can be considered as the argument supporting behavior noted by Forsmann et al.[57]. Pressure evolutions of primary/structural relaxation time and conductivity seem to follow the linear dependence, what can support the simple portrayal via the Barus equation[50], the pressure equivalent of the basic Arrhenius dependence.

Notwithstanding, results of the derivative based and distortions-sensitive analysis presented in Fig. 4 reveal notable distortions from the simple Barus-type evolution. Fig. 4 focuses on the evolution of the apparent activation volume, obtained via:[45, 60]

$$V'(P) = \frac{V(P)}{RT} = \frac{d \ln \tau(P)}{dP} \qquad (13)$$

For the Barus-type behavior the horizontal line $V'(P) = V = const$ should appeared. The obtained pattern is notably different.

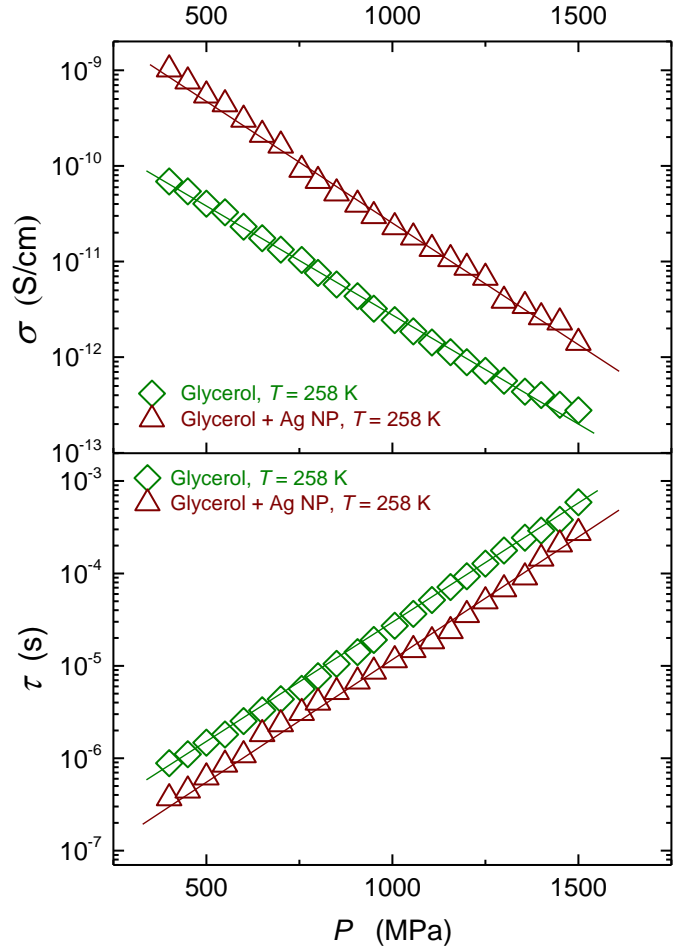

**FIG. 3** *The pressure evolution of the primary relaxation time (τ) and DC conductivity (σ) for pure glycerol and glycerol + Ag NP composite. Lines are guides for eyes.*

The evolution of the apparent free volume for the pure glycerol and the nanocolloid glycerol + AgNP are strongly different. The latter shows "singularity" near $P \approx 0.65 GPa$ and the crossover to clearly super-Barus (SB) behavior for $P > 1.2 GPa$. It is worth stressing that the apparent activation volume is directly related to the pressure counterpart of the steepness index and fragility:[45, 60]

$$m_T(P) = \left[\frac{\log_{10}\tau}{d\Delta P/\Delta P_g}\right]^{T=const} \quad (9) \qquad \text{one then:} \qquad V_a' = \frac{RT_g}{\ln 10(\Delta P_g)}m_T \quad (14)$$

where $\Delta P_g = P_g - \pi$, where $\pi$ denotes the terminal $P_g(T \to 0) = \pi$ in the negative pressures domain.

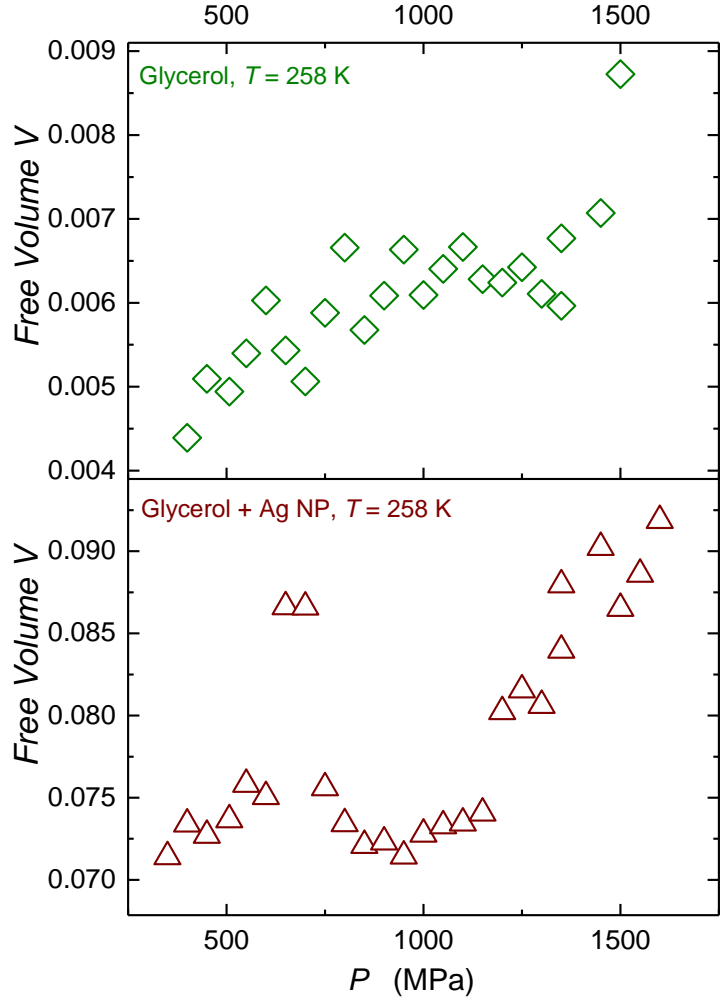

**FIG. 4** *The complex SB/SA evolution of the normalized free volume (V) in pure glycerol and glycerol + AgNP composite at temperature T=258 K*

For $P = P_g$ eq. (14) defines the pressure counterpart of the temperature fragility index $m$, defined above.[45,60] It is notable, that the "temperature path" fragility $m = m_T(T_g)$ and the "pressure path" fragility $m_P(P_g)$ defined via eq. (14) can be considered as "parallel parameters". For instance, the minimal values of the fragilities related to the simply Arrhenius or Barus behavior are given as: $m_{\min} = \log_{10}\tau(T_g) - \log_{10}\tau_o^{VFT} = 2 - \log_{10}\tau_0^{VFT} \approx 16$, assuming the average value for the prefactor $\tau_0^{VFT} \approx 16$.[3] Approximately the same value can be assumed for the prefactor in the pressure related eq. (2), and then $m_{P-\min} = \log_{10}\tau(T_g) - \log_{10}\tau_o^{P} = 2 - \log_{10}\tau_0^{P} \approx 16$.[45, 60] Such equivalence does not occur for the often used definition $m_T(P_g) = [d\log_{10}\tau(T)/(P/P_g)]_{P=P_g}^{T-T_g}$,[3,16] which recalls the pressure counterpart of the VFT equation from ref.[47] $\tau(P) = \tau_0^P \exp(D_P P/(P_0 - P))$, where the prefactor $\tau_0^P$ can range from seconds to picoseconds.[45,58] In each case the existence of the negative pressures domain was neglected and $P = 0$ was takes as the reference.

Experimental data shown in Figs. 5 and 6 yield the possibility of the insight into the coupling/decoupling between translational and orientational degrees of freedom for compressed glycerol and its Ag-based nanocollid / nanocomposite. When increasing pressure, first the dynamics associated with fractional exponent $S \approx 1$ takes place. This indicates the coupling between orientational and translational processes.

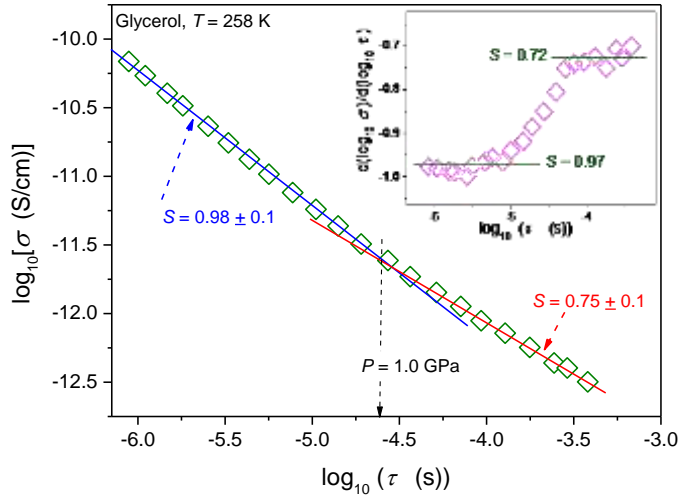

**FIG. 5** *Test of the fractional Debye-Stokes-Einstein behavior in pressurized glycerol at T=258 K. The inset shows results of the derivative-based, distortions sensitive analysis of data from the main part of the plot.*

However when entering the GPa domain crossovers to the clearly decoupled domains occur. For pure glycerol this is associated with $P \approx 1 GPa$ and $S \approx 0.75$ and for the nanocolloid with $P \approx 1.2 GPa$ and $S \approx 0.55$, probably the smallest value reported so far.[3, 14, 16, 19-31]

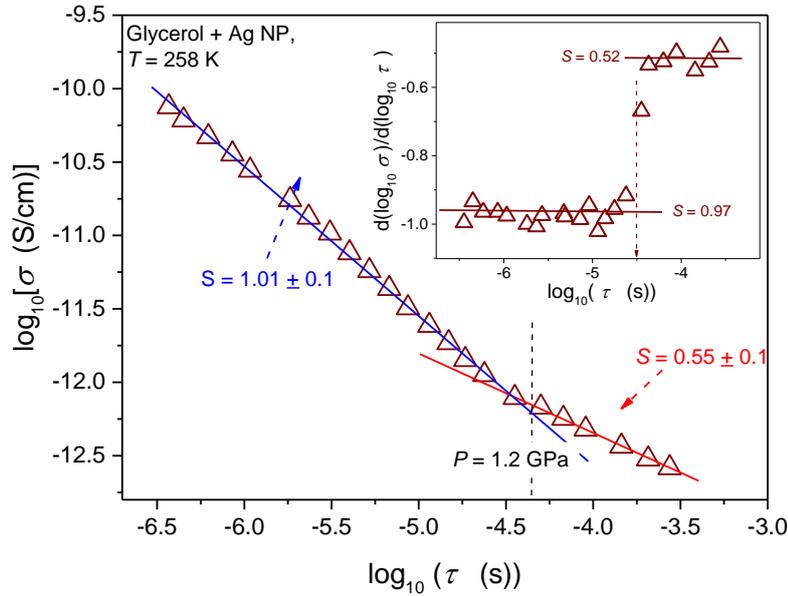

**FIG. 6**  *Test of the fractional Debye-Stokes-Einstein behavior in pressurized nanocolloid (glycerol + Ag NP) at T=258 K. The inset shows results of the derivative-based, distortions sensitive analysis of data from the main part of the plot.*

Results of the derivative-based analysis in insets in Figs. 5 and 6 confirm the picture from main parts of these plots, but with slightly lower values of FDSE exponents. It also shows that for glycerol the transformation from coupled (DSE) to the decoupled (FDSE) domain is stretched over one decade in time or $\Delta P \approx 0.3 GPa$. For the nanocolloid the crossover is sharp and "strong", i.e. associated with the transformation from $S \approx 1$ to $S \approx 0.5$.

**Conclusions**

Glycerol is a versatile compound due to its enormous significance in a variety of applications ranging from biotechnology to pharmacy, cosmetics, "*green and biodegradable*" plastics, textiles and foodstuffs industries.[61-64] The practical significance of results presented extents also for the nanocolloid due to the the great antimicrobial activity of Ag nanoparticles[65] and the fact that glycerol based Ag-nanocolloids can serve as an important carrier platform for practical implementations recalled above.

From the fundamental point of view glycerol has a simple molecular structure, large permanent dipole moment and the relatively small electric conductivity, what coincides with preferred features for the broad band dielectric spectroscopy (BDS) monitoring.[15] It can be also very easily supercooled. All these caused that glycerol has gained the position of a model "classical" system in glass transition studies.

This report shows that for ultraviscous glycerol the translational – orientational decoupling tested via $\sigma(T)$ vs. $\tau(T)$ seems to be almost absent, i.e. the FDSE exponent $S \approx 1$. Such value is characteristic for the non-ultraviscous region for $T > T_B$ ($P < P_B$). It is noteworthy that tests based on $D(T)$ and $\eta(T)$ experiments indicated for glycerol clearly decoupled behavior with the FDSE exponent $\zeta \approx 0.85$.[11] The addition of Ag nanoparticles to glycerol, yielding stable nanocolloid, qualitatively changes the "ultraviscous dynamics". For temperature tests under atmospheric pressure there are two domains: "coupled" (with $S \approx 1$) and decoupled (with $S \approx 0.91$) near the glass temperature.

Results presented below addressed also the fundamental question of the validity of the DS equation $\tau \propto \eta/T$ vs. the Maxwell relation $\tau \propto \eta$.[18, 39] The clear prevalence of the latter, with $G_\infty \approx const$, for ultraviscous glycerol is indicated. It is notable that in complex and ultraviscous liquids for portraying the evolution of viscosity and relaxation time VFT eq. (1) is most often used. For the electric conductivity its corrected version $\sigma^{-1}(T) = (\sigma_0/T)\exp[D^\sigma/(T - T_0)]$ is preferred.[3, 15] Its implementation can be justified via the general validity of DS eq. (3). However, for the ultraviscous domain it should be at least supplemented by the FDSE exponent resulted from the fractional form of eq. (5)

$\sigma \tau^S / T = const$. This report and evidence from refs.[16, 19-33] indicate the validity of eq. (7) $\sigma(T)[\tau(T)]^S = const$. All these lead to the relation: $\sigma^{-1}(T) = \sigma_0 \exp[SD_T T_0 / (T - T_0)]$, with $D_T^\sigma = SD_T T_0$. The implementation of consequences of FDSE behavior for systems where the critical-like parameterization of $\tau(T)$ and $\sigma(T)$ is optimal is given in ref.[35].

The "decoupling dynamics" becomes even more complex for superpressed systems. In glycerol the unusual and does not reported so far crossover between the "coupled" ($S \approx 1$) to notably "decoupled" ($S \approx 0.72$) domains appers. The transition between these regions is stretched and covers up to $\Delta P \approx 0.3 GPa$. For the nanocolloid such transition takes place at the well-defined pressure $P = 1.25 GPa$ and it is associated with extraordinary small value of the FDSE exponent $S \approx 0.5$. This can indicate the possible the fluid – fluid transition.

The simple analysis presented above indicates that the FDSE behavior for $D(T)$ vs. $\eta(T)$, $D(T)$ vs. $\tau(T)$ (eq. (1) and $\sigma(T)$ and $\tau(T)$ (eqs. (7) and (11)) can be correlated i.e. exponents $\zeta = S$. For the FDSE exponent $\zeta$ Mallamace et al.[11] suggested a universal value $\zeta = 0.85$ for arbitrary ultraviscous liquid, basing on the empirical analysis for few tens of ultraviscous low molecular weight liquids (including glycerol).

Concluding, results presented show new puzzling dynamics-related artefacts in glycerol, one of the most "classical" ultraviscous glass forming liquids. They are particularly notable for glycerol with the addition of solid Ag nanoparticles. This can indicate that such nanocolloidal systems can yield new and interesting class of glass formers. The application of the high hydrostatic pressure can lead to a unique fluid-fluid transition in the GPa domain. Results presented are also worth stressing when taking into account the postulated link of FDSE exponent as one of metrics of dynamic heterogenic, still mysterious beings hypothetically responsible for universalities emerging in the previtreous domain.

.

**Acknowledgements**

This research was supported by the National Science Centre (Poland) via grant 2011/01/B/NZ9/02537.